\documentclass[runningheads]{llncs}
\usepackage{graphicx}
\usepackage{diagbox}
\usepackage{xcolor}
\usepackage{makeidx}  
\usepackage[colorlinks,linkcolor=blue]{hyperref}
\usepackage{amsmath, amsfonts}
\usepackage{booktabs}
\usepackage[misc]{ifsym} 
\usepackage{bbding}
\usepackage{amsfonts}
\usepackage{multirow}
\usepackage{xspace}
\usepackage[colorlinks,linkcolor=blue]{hyperref}

\begin{document}
\title{Explainable and Controllable Motion Curve Guided Cardiac Ultrasound Video Generation}
\author{}
\institute{}
\titlerunning{Cardiac Ultrasound Video Generation}


\author{Junxuan Yu\inst{1,2,3}\thanks{Junxuan Yu and Rusi Chen contribute equally to this work.} \and Rusi Chen\inst{1,2,3\star} \and Yongsong Zhou\inst{1,2,3} \and Yanlin Chen\inst{1,2,3} \and Yaofei Duan\inst{4} \and Yuhao Huang\inst{1,2,3} \and Han Zhou\inst{1,2,3,5} \and Tan Tao\inst{4} \and Xin Yang\inst{1,2,3}\textsuperscript{(\Letter)} \and Dong Ni\inst{1,2,3}\textsuperscript{(\Letter)}} 

\institute{
\textsuperscript{$1$}National-Regional Key Technology Engineering Laboratory for Medical Ultrasound, School of Biomedical Engineering, Medical School,
Shenzhen University, China\\
\email{xinyang@szu.edu.cn}; \email{nidong@szu.edu.cn} \\
\textsuperscript{$2$}Medical Ultrasound Image Computing (MUSIC) Lab, Shenzhen University, China\\
\textsuperscript{$3$}Marshall Laboratory of Biomedical Engineering, Shenzhen University, China\\
\textsuperscript{$4$}Faculty of Applied Sciences Macao Polytechnic University, Macao, China \\
\textsuperscript{$5$}Shenzhen RayShape Medical Technology Co., Ltd, China\\
}

\authorrunning{J.Yu and R.Chen et al.}

\maketitle

\begin{abstract}
Echocardiography video is a primary modality for diagnosing heart diseases, but the limited data poses challenges for both clinical teaching and machine learning training.
Recently, video generative models have emerged as a promising strategy to alleviate this issue.
However, previous methods often relied on holistic conditions during generation, hindering the flexible movement control over specific cardiac structures.
In this context, we propose an explainable and controllable method for echocardiography video generation, taking an initial frame and a motion curve as guidance. 
Our contributions are three-fold.
\textbf{First}, we extract motion information from each heart substructure to construct motion curves, enabling the diffusion model to synthesize customized echocardiography videos by modifying these curves.
\textbf{Second}, we propose the structure-to-motion alignment module, which can map semantic features onto motion curves across cardiac structures.
\textbf{Third}, The position-aware attention mechanism is designed to enhance video consistency utilizing Gaussian masks with structural position information.
Extensive experiments on three echocardiography datasets show that our method outperforms others regarding fidelity and consistency. The full code will be released at \href{https://github.com/mlmi-2024-72/ECM}{https://github.com/mlmi-2024-72/ECM}.

\end{abstract}

\section{Introduction}
Echocardiography is a primary method that relies on dynamic video to obtain structural information for clinical diagnoses \cite{1-2}. 
However, training radiologists with diagnostic skills and establishing machine learning models both suffer from limitations on video resources.
Recently, video generation models have demonstrated a promising ability to solve this problem, owing to their powerful capability in modeling data distribution~\cite{xing2023survey,wu2023tune}.
Several studies about video generation based on specific conditions have been investigated.
The typical ones relied on canny edges or depth maps~\cite{chen2023control,wang2024videocomposer} extracted from additional videos as conditions. Nevertheless, these conditions are non-editable and lack motion-driving information.
In contrast, Shi et al. \cite{shi2024motion} proposed the Motion-I2V framework to predict dense optical flow and guide video generation, maintaining both spatial and motion consistency.
Wang et al. \cite{wang2024boximator} then employed a self-tracking training method.
Specifically, they specified the box positions of the first and last frames along with motion paths, to control the movement of objects.
However, these paths are basic and coarse, inadequate for ultrasound video synthesis that demands an accurate representation of intricate structure motions.

\begin{figure}[!t]
	\begin{center}
        \includegraphics[width=1\columnwidth]{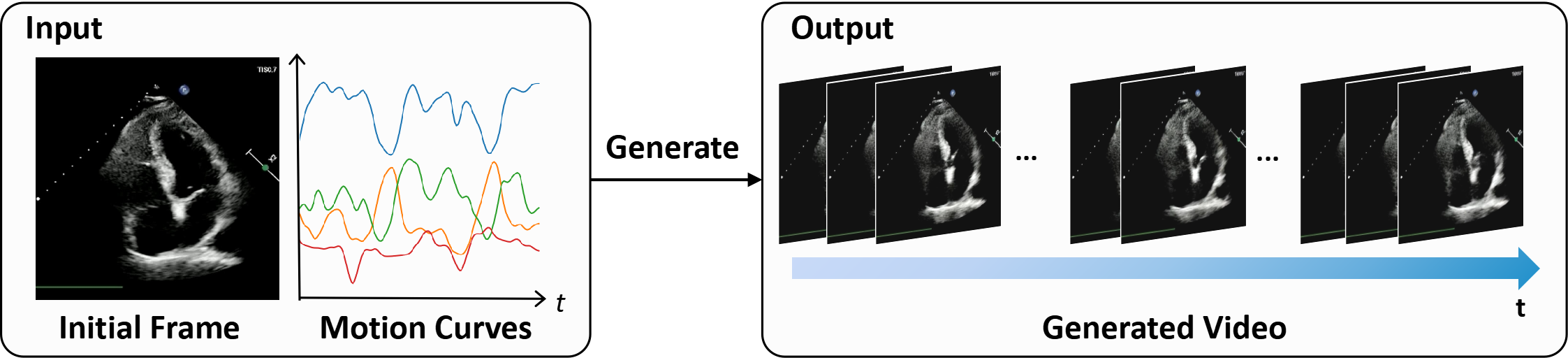}
	\end{center}
	\caption{Workflow of ECM. Input: an initial frame and motion curves of each cardiac structure. Output: a generated echocardiography video.}
	\label{task}
\vspace{-0.5cm}
\end{figure}

In the field of echocardiography video generation, 
Zhou et al. \cite{zhou2023onuvs} proposed the OnUVS framework to synthesize ultrasound videos by animating source images and leveraging motion information from driving video. 
Nevertheless, OnUVS faced challenges due to the anatomical structure gap between the source image and the driving video, making accurate motion control difficult.
In addition, some studies have mined the structural information of the heart to guide the movement of videos.
For instance, Reynaud et al.~\cite{reynaud2022d} developed a Generative Adversarial Network (GAN) capable of generating echocardiography videos corresponding to the left ventricular ejection fractions (LVEFs).
They further employed a cascade video diffusion model conditioned on randomly sampled frames, enhancing the synthesis quality of echocardiography video~\cite{reynaud2023feature}.
Van et al. \cite{van2024echocardiography} utilized segmentation masks of end-diastolic (ED) frames as a condition to generate four-chamber heart videos.
However, both LVEFs and ED masks are relatively sparse conditions, leading to an imbalance of information between intricate motions and limited conditions. 
This sparsity poses challenges for effectively controlling fine-grained cardiac structures, thus limiting the ability to capture the full complexity of heart movements and dynamics.

To address the above issues, we propose an explainable and controllable motion curve guided video diffusion model (\textbf{ECM}) that can synthesize video guided by the initial frame and motion curves (Fig. \ref{task}). 
Our contributions are threefold: 
(1) We innovatively mine the motion of the echocardiography video to obtain motion curves, which fully reflect the movement of each cardiac structure. This easily controlled approach enables the customization of videos through the modification (scaling and replacing) of the initial motion curve.
(2) As the curve lacks category information of the structure, we propose a Structure-to-Motion alignment mechanism. This mechanism extracts the semantic features of each cardiac structure and maps them with motion curve features, aiming to align visual and motion information effectively.
(3) We design position-aware attention masks based on the position of the cardiac structure movement, effectively enhancing the motion consistency of each structure.
To the best of our knowledge, ECM is the first study to apply cardiac motion curve guidance in echocardiography video generation. Extensive experimental results show that the proposed ECM is a flexible, controllable, and reliable method.

\section{Methodology}
\begin{figure}[!t]
	\begin{center}
			\includegraphics[width=1\columnwidth]{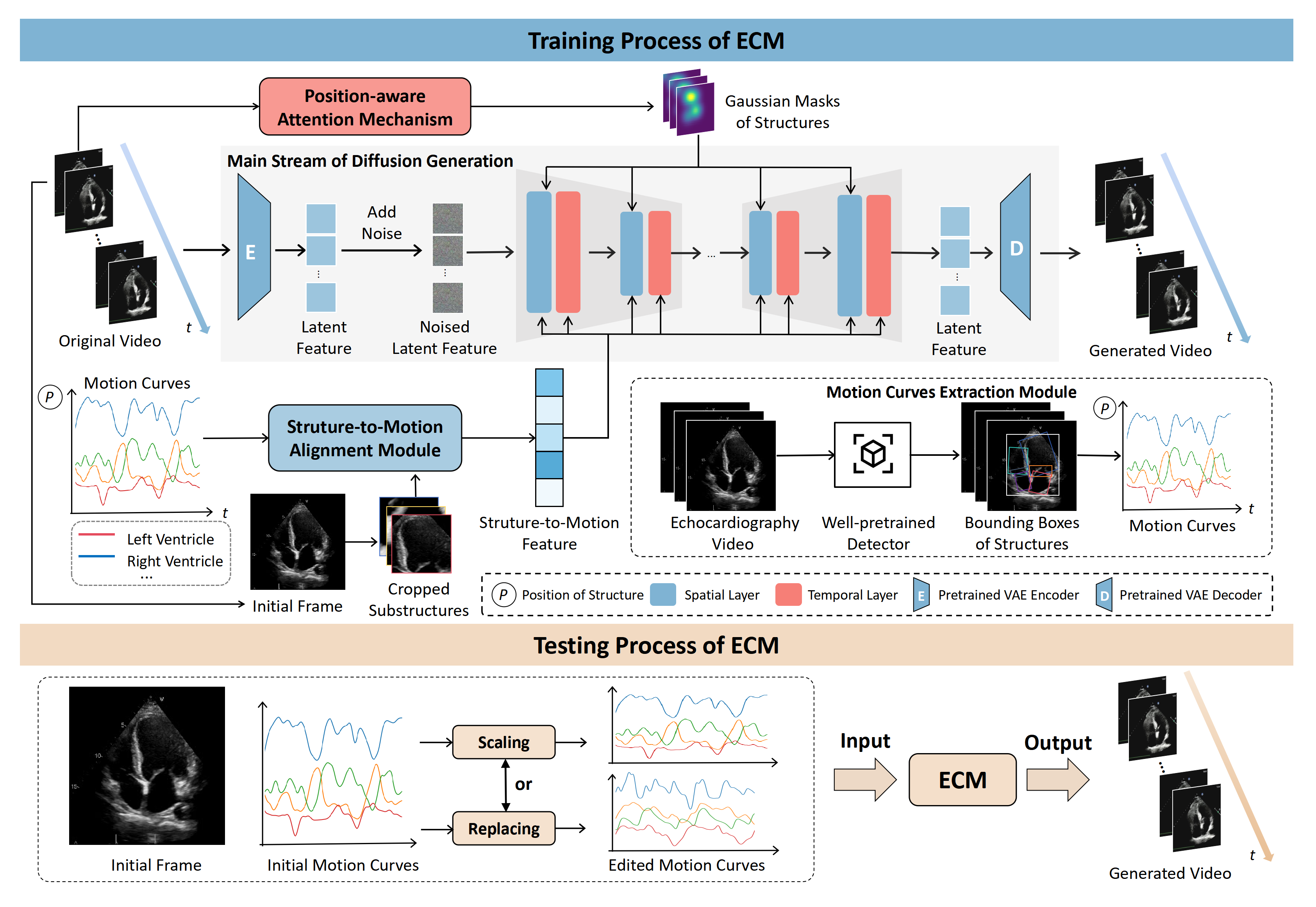}
	\end{center}
	\caption{The overall pipeline of the proposed ECM. 
 }
	\label{fig:2}
\end{figure}

Fig.~\ref{fig:2} shows the framework of ECM. During the training process, ECM takes an original video as input. 
A pretrained variational auto-encoder (VAE) from Stable Diffusion (SD) \cite{rombach2022high} is then utilized to downsample the video into latent features. 
By gradually adding noise to the latent and then learning to denoise it, the model can obtain the generated video with a pretrained VAE decoder. 
Remarkably, we develop the structure-to-motion alignment module to match cardiac structures with motion, yielding aligned features that condition the ECM model. 
Additionally, we employ Gaussian masks for each structure using a Position-aware Attention Mechanism, incorporating these into the spatial layers of the diffusion model. 
During the testing process, ECM generates echocardiography videos using an initial frame and motion curves as input. Users can customize the videos by replacing the initial frame or adjusting (scaling or replacing) the motion curves, highlighting the controllability and interpretability of our method.

\subsection{Extraction of Motion Curves}\label{subsec:motion_curve}
Previous studies~\cite{reynaud2022d,reynaud2023feature} faced challenges in controlling the motion of specific cardiac substructures since they relied solely on the single sparse condition (LVEFs), while our approach aims to provide fine-grained control over the motion of each cardiac substructure.
Therefore, we extract motion curves for the key structures (e.g., Left Ventricle, Left Atrium, Mitral Valve, etc.) as conditions.

As illustrated in Fig. \ref{fig:2}, the process of extracting motion curves is as follows: 
\textbf{(a)} We employ a well-trained anatomy detector (average accuracy=85\%) to identify each substructure in each frame of the echocardiogram video. 
\textbf{(b)} Subsequently, we utilize the pixel coordinates of each substructure's bounding box (bbox) as the basis for encoding the motion curves, which form the basis for encoding the motion curves, represented as 
$ f_{c}^{m} \in \mathbb{R}^{B \times N \times C \times ( 4 \times 2)} $,
where N represents the number of frames and C represents the categories of the substructure. Notably, any missed detected structures would be treated as learnable parameters, initialized by the network.
\textbf{(c)} Since cardiac motion is periodic, we employ Fourier transformation (\textit{FT}) to transform the pixel coordinates into a high-dimensional feature representation, denoted as 
$ f_{c}^{m} \in \mathbb{R}^{B \times N \times C \times E} $, 
where \( E \) represents the dimensionality of the features obtained from \textit{FT}. 
\textbf{(d)} Finally, the motion embedding is passed through several multi-layer perceptron (MLP) layers. Overall, these motion curve features can be formulated as $ f_{c}^{m} \in \mathbb{R}^{B \times N \times C \times 1024} $.

\subsection{Structure-to-Motion Alignment Module}

Although the motion curves of the echocardiography videos are captured, it is challenging for the model to distinguish the relationship between the motion curves and the semantic information of each substructure.
Recently, GLIGEN~\cite{li2023gligen} has demonstrated its effectiveness in combining caption and bbox information, enhancing visual-language understanding to enable fine-grained control over specific objects in natural images.
However, our preliminary experiments suggested that GLIGEN struggles to effectively represent and interpret texts related to cardiac structures in echocardiography.

\begin{figure}[!h]
	\begin{center}
        \includegraphics[width=0.98\columnwidth]{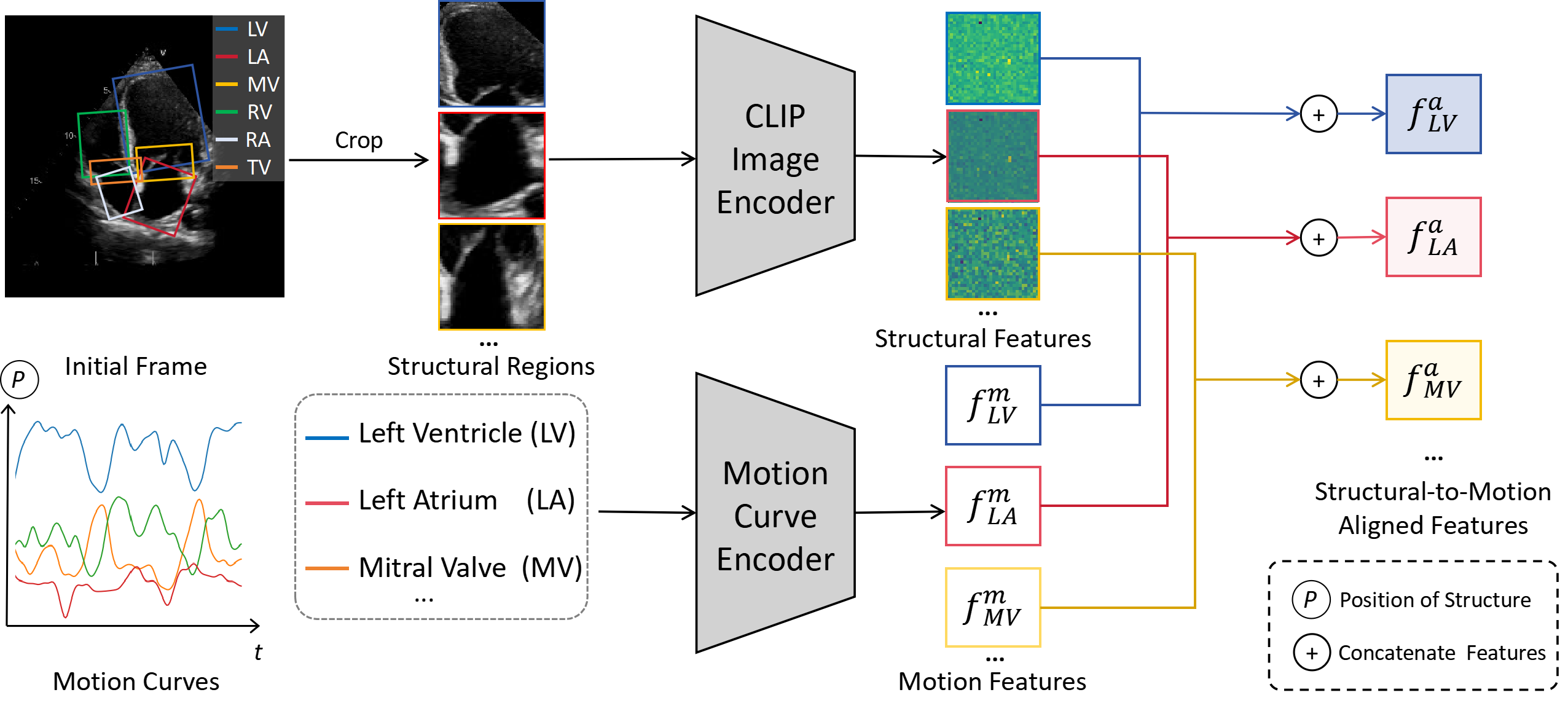}
	\end{center}
	\caption{Illustration of structure-to-motion alignment module.}
	\label{fig:3}
\end{figure}

Innovatively, we replace texts with the cardiac structure features, denoted as $f_{i}^{s}$. 
As shown in Fig. \ref{fig:2}, to obtain $f_{i}^{s}$, 
the regions of interest (ROI) that correspond to the cardiac substructures in the initial frame are cropped by the well-pretrained detector as mentioned above. 
Then, a pretrained CLIP image encoder \cite{radford2021learning} is utilized to transform the ROIs into structural embedding features.
Then, these features pass through several MLP layers,
resulting in an output denoted as $ f_{c}^{s} \in \mathbb{R}^{B \times N \times C \times 1024} $, which has the same shape with the extracted motion curve features $f_{i}^{m}$ mentioned in Sec.~\ref{subsec:motion_curve}.
Consequently, structural features and their corresponding motion curve features from the same category are concatenated to create aligned features. The aligned features are formulated as:
\begin{equation}
    {F_{c}^{a}} = {Concat(f_{c}^{s}, f_{c}^{m})}.
\end{equation}

Note that any undetected structures are replaced with general features from the dataset.
Next, the concatenated features pass through an additional MLP layer for further integration.
Finally, to introduce the aligned motion curve features to guide the generation of echocardiography video, we then mapped 
\({F_{c}^{a}}\) to the intermediate spatial layers of the UNet\cite{ronneberger2015u} via a cross-attention layer implementing cross attention, formulated as:
\begin{equation}
\text{CrossAtten}(Q, K, V) = \text{softmax} \left( \frac{QK^T}{\sqrt{d}} \right) \cdot V,
\end{equation}
where $Q$, $K$ and $V$ represent the query, key and value respectively in the attention mechanism. Here, we regard noise latent feature $ F_{l}^{m} \in \mathbb{R}^{B \times 64 \times 64} $ as query and the structure-to-motion embedding features as key and value.

\subsection{Position-aware Attention Mechanism}
Enhancing the consistency of cardiac motion is crucial in the task of cardiac video generation. 
To address this, we aim to inject the positional information of the cardiac structure into the cross-attention mechanism. Specifically, we design Gaussian masks that are generated based on the position of the cardiac structures. 
The Gaussian masks are defined as:
\begin{equation}
M_{g}(x, y) = \frac{1}{2\pi \sigma^2} \exp\left(-\frac{(x - \mu_x)^2 + (y - \mu_y)^2}{2\sigma^2}\right),
\end{equation}
where \(x\) and \(y\) represent the spatial positions of the mask.
\(\mu_x\) and \(\mu_y\) denote the center positions of the Gaussian distribution, which are the coordinates of the four corner points of the detected bbox. 
\(\sigma\) is the standard deviation to control the distribution width. 
Here, we set \(\sigma\) to 10 pixels as default.
Subsequently, the Gaussian masks are resized to match the dimensions of the latent feature map. 
The Gaussian-weighted cross-attention mechanism is as follows:
\begin{equation}
\text{CrossAttenMask}(Q, K, V, M_g) = \text{softmax} \left( \frac{QK^T}{\sqrt{d_k}} {\odot}M_g \right) V,
\end{equation}
where $\odot$ is the element-wise multiplication. Overall, this position-aware attention mechanism effectively integrates positional information of cardiac structures, enhancing the consistency and realism of generated cardiac motion.

\section{Experiments}

\textbf{Datasets and Implementations.} 
To assess the performance of ECM, we gathered data from three sources: two private dataset from multiple hospitals and the publicly available dataset named EchoNet-Dynamic~\cite{ouyang2020video}.
The in-house dataset includes 144 apical four-chamber (A4C) and 100 apical two-chamber (A2C) heart videos, and the public one comprises 10,030 labeled A4C echocardiography videos.
For both datasets, the videos were randomly split into training (90\%) and testing (10\%) sets. During training, 12-frame clips were randomly sampled from each video, with a sampling interval ranging from 1 to 4 frames. 
For testing, videos were truncated according to different sampling intervals. 
The input videos from the private datasets were resized to 256x256 pixels, while those from the EchoNet-Dynamic dataset were kept at their original resolution of 112x112 pixels.
All methods were implemented in PyTorch using an NVIDIA RTX 4090 GPU under same settings. The Adam optimizer was used with a learning rate of 5e-3 and 60,000 training steps.

\textbf{Evaluation Metrics.}
Our evaluation metrics cover both image-level and video-level evaluation.
The image-level assessment includes Structural Similarity Index (SSIM) \cite{SSIM}, Mean Absolute Error (MAE) \cite{MAE}, Peak Signal-to-Noise Ratio (PSNR) \cite{PSNR}, Fréchet Inception Distance (FID) \cite{gans} and Learned Perceptual Image Patch Similarity (LPIPS) \cite{29}.
For video-level assessment, we only considered the commonly-used Fréchet Video Distance (FVD)~\cite{fvd}.
Notably, to assess the consistency of cardiac structures between the generated and target videos, we introduce a new indicator calculated by the Intersection over Union (IoU) between the bboxes of the original and synthesized videos.

\begin{table}[!h]
\setlength{\tabcolsep}{1pt}
\caption{\textbf{Comparison of ECM with other generative methods on A4C and EchoNet-Dynamic dataset.} \textcolor{blue}{blue} emphasizes the optimal results. \textbf{Bbox} refers to the bounding box of structure, \textbf{Text} means a fixed prompt as \textit{`This is an echocardiography video'}, and \textbf{Canny} represents the Canny edge map. \textbf{IF} represents the initial frame.}
\resizebox{\textwidth}{!}{
\begin{tabular}{c|l|c|cccccc}
\toprule[1.2pt] 
Dataset & Method           & Condition       & SSIM↑ & MAE↓  & PSNR↑  & FID↓                                  & Lpips↓ & FVD↓                                  \\ \hline
\multirow{2}{*}{} &
  SEG Diffusion  \cite{olive2023synthetic}  & /               & 0.030 & 0.200 & 11.300 & /                                     & 0.450  & 3107.70                               \\
  & Stable Diffusion \cite{rombach2022high}& Bbox/Text       & 0.660 & 0.050 & 18.700 & 66.890                                & 0.170  & 792.19                                \\
  & Stable Diffusion \cite{rombach2022high}& Bbox/Text/Canny & 0.640 & 0.050 & 18.740 & \textcolor{blue}{53.200}                                & 0.360  & 1037.95                               \\
  & 3D ControlNet \cite{3DControlNet}   & Canny           & 0.421 & 0.220 & 8.012  & /                                     & 0.583  & 1540.52                               \\
\multirow{-5}{*}{A4C} &
  \textbf{ECM(Ours)} &
  IF/Motion Curves &
   \textcolor{blue}{0.719} &
  \textcolor{blue}{0.038} &
  \textcolor{blue}{21.762} &
  79.14 &
  \textcolor{blue}{0.114} &
  \textcolor{blue}{189.27} \\ \hline
\multirow{2}{*}{EchoNet-Dynamic} &
  EchoDiffusion  \cite{reynaud2023feature}  & IF/LVEF      & 0.530 & 9.650 & -      & \textcolor{blue}{12.30} & 0.210  & \textcolor{blue}{60.50} \\
  & \textbf{ECM(Ours)} &
  IF/Motion Curves &
  \textcolor{blue}{0.611} &
  \textcolor{blue}{0.057} &
  \textcolor{blue}{19.450} &
  36.80 &
  \textcolor{blue}{0.118} &
  109.86 \\ 
\toprule[1.2pt] 
\end{tabular}}
\label{Comparison_result}
\end{table}

\textbf{Method Comparison.}
The quantitative comparison of ECM and other methods is reported in Table \ref{Comparison_result}.
The ECM model achieves the best performance across all metrics, indicating superior image quality and high fidelity compared to other methods.
It can be seen that the SEG Diffusion method, generating echocardiography videos without any control conditions, achieves poor quality.
Notably, ECM model markedly outperforms the SD method under bbox/text conditions, achieving 67.8\%$\downarrow$ in FVD. 
This demonstrates that synthesizing 2D images with rich conditions and stitching them into videos does not yield satisfactory outcomes. 
The integration of a position-aware attention mechanism significantly contributes to our superior performance and enhanced video consistency.
Furthermore, the ECM model shows better performance than 3D ControlNet. 
This disparity arises from the fact that natural images typically have clearly defined control conditions, whereas the motion in echocardiography videos is inherently more complex.
Similarly, for the EchoNet-Dynamic dataset, guided by motion curves, our proposed ECM outperforms the strong competitor~\cite{reynaud2023feature} across most metrics, indicating its superior generation capacity.

\begin{table}[!h]
\caption{\textbf{Ablation results of ECM model on A4C and A2C datasets.} Note that \textbf{Base} refers to training generation model without any condition. \textbf{Text} means replacing the motion curves with a fixed prompt.
\textbf{S2M} and \textbf{Att} represent structrue-to-motion alignment module and position-aware Attention Mechanism, respectively.
}
\label{quantitative_results}
\resizebox{\textwidth}{!}{
\centering
\begin{tabular}{clccccccc}
\toprule[1.2pt] 
&
   &
  \multicolumn{5}{c}{Image-level Metrics} &
  \multicolumn{2}{c}{Video-level Metrics} \\ \hline
\multicolumn{1}{c|}{Dataset} &
  \multicolumn{1}{l|}{Condition} &
  SSIM↑ &
  MAE↓ &
  PSNR↑ &
  FID↓ &
  \multicolumn{1}{c|}{Lpips↓} &
  FVD↓ &
  IoU↑ \\ \hline
\multicolumn{1}{c|}{\multirow{5}{*}{A4C}} & \multicolumn{1}{l|}{Base+Text}          & 0.714 & 0.040 & 21.296 & 75.85 & \multicolumn{1}{c|}{0.123} & 307.50 & 0.691 \\
\multicolumn{1}{c|}{} &
  \multicolumn{1}{l|}{Base+Motion} &
  0.718 &
  0.039 &
  21.565 &
  79.91 &
  \multicolumn{1}{c|}{\textcolor{blue}{0.112}} &
  259.77 &
  0.748 \\
\multicolumn{1}{c|}{} &
  \multicolumn{1}{l|}{Base+Motion+S2M} &
  0.710 &
  0.042 &
  21.258 &
  \textcolor{blue}{66.56} &
  \multicolumn{1}{c|}{0.121} &
  228.23 &
  0.717 \\
\multicolumn{1}{c|}{} &
  \multicolumn{1}{l|}{Base+Motion+Att} &
  0.718 &
  0.039 &
  21.583 &
  78.20 &
  \multicolumn{1}{c|}{0.115} &
  203.83 &
  0.751 \\
\multicolumn{1}{c|}{}                     & \multicolumn{1}{l|}{\textbf{ECM(Ours)}} & \textcolor{blue}{0.719} & \textcolor{blue}{0.038} & \textcolor{blue}{21.762} & 79.14 & \multicolumn{1}{c|}{0.114} & \textcolor{blue}{189.27} & \textcolor{blue}{0.766} \\ \hline
\multicolumn{1}{c|}{\multirow{4}{*}{A2C}} & \multicolumn{1}{l|}{Base+Motion}        & 0.694 & 0.039 & 22.480 & 40.12 & \multicolumn{1}{c|}{\textcolor{blue}{0.093}} & 269.23 & 0.804 \\
\multicolumn{1}{c|}{} &
  \multicolumn{1}{l|}{Base+Motion+S2M} &
  0.692 &
  0.037 &
  22.056 &
  37.83 &
  \multicolumn{1}{c|}{0.099} &
  232.43 &
  0.794 \\
\multicolumn{1}{c|}{} &
  \multicolumn{1}{l|}{Base+Motion+Att} & 0.701
   & \textcolor{blue}{0.036} 
   & \textcolor{blue}{22.487}
   & 36.81
   & 
  \multicolumn{1}{c|}{0.098} & 286.87
   & 0.819
   \\
\multicolumn{1}{c|}{} &
  \multicolumn{1}{l|}{\textbf{ECM(Ours)}} & \textcolor{blue}{0.709}
   & \textcolor{blue}{0.036} 
   & 22.349
   & \textcolor{blue}{35.73}
   & 
  \multicolumn{1}{c|}{0.097} & \textcolor{blue}{208.12}
   & \textcolor{blue}{0.828}
   \\
\midrule[1.2pt] 
\end{tabular}%
}
\end{table}

\begin{figure}[!h]
	\begin{center}
			\includegraphics[width=1.0\columnwidth]{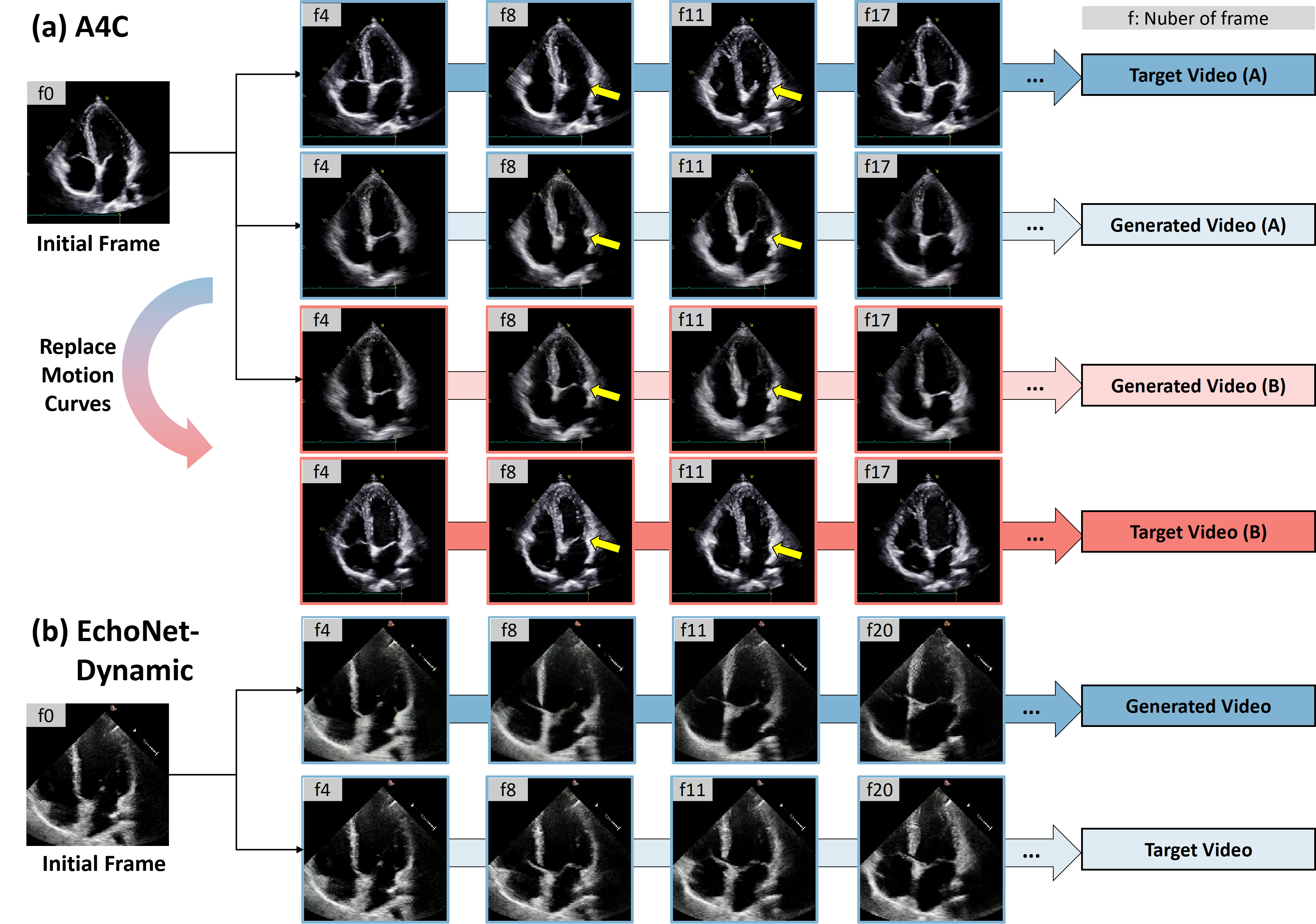}
	\end{center}
	\caption{Visualization results of generated videos in two datasets.
 }
	\label{vis_result}
\end{figure}

\textbf{Ablation Study.}
We conducted the ablation study to test the contribution of each component in Table \ref{quantitative_results}. 
The ECM model consistently outperforms others across most metrics for both the A4C and A2C datasets, achieving outstanding FVD scores of 189.27 and 208.12, and impressive IoU scores of 0.766 and 0.828, respectively.
This highlights that ECM effectively enhances video generation consistency while maintaining strong image quality.
Additionally, incorporating motion curves instead of traditional textual control leads to notable improvements. 
Specifically, using motion curves, the FVD decreased from 307.50 to 259.77, and the IoU increased from 0.691 to 0.748.
This indicates that motion curves provide precise control over the motion in echocardiography videos.
It can also be observed that the Structure-to-Motion alignment module (`+S2M') and the position-aware attention mechanism (`+Att') enhance image quality and video motion consistency, respectively. Consequently, the final ECM model demonstrates excellent performance at both the image and video levels.

\textbf{Qualitative Results.} 
Fig. \ref{vis_result} demonstrates that ECM generates videos closely matching the target in both the A4C and EchoNet-Dynamic datasets by inputting the initial frame and motion curves. It effectively mimics heart structure motions such as chamber dilation, contraction, and diastolic opening and systolic closing
Furthermore, Fig. \ref{vis_result} (a) illustrates ECM's controllability. Starting from the same initial frame, the second-row video uses the original motion curves from target video A, while the third-row video uses replaced curves from target video B. The mitral valve's systole and diastole differ after replacing the motion curves (as shown by yellow arrows), indicating ECM's effective control over motion curves.
More visualization results, including generated videos with replaced and scaled motion curves, can be found at \href{http://106.53.160.189:8000/}{anonymous link}.

\section{Conclusion}
In this study, We have presented ECM for generating echocardiography videos guided by motion curves, which reflect the movement of each cardiac structure. ECM enables to customize the generated videos by adjusting (scaling and replacing) the initial motion curve. Besides, to link the structure features with corresponding movement information, we propose the structure-to-motion alignment mechanism. Moreover, attention masks based on the position of the anatomical structures are introduced to enhance the motion consistency of each structure. Overall, our proposed ECM achieves state-of-the-art performance for generating echocardiography videos in terms of fidelity and consistency.

\begin{credits}
\subsubsection{\ackname}
This work was supported by the grant from National Natural Science Foundation of China (12326619, 62101343, 62171290), Science and Technology Planning Project of Guangdong Province (2023A0505020002), Science and Technology Development Fund of Macao (0021/2022/AGJ), and Shenzhen-Hong Kong Joint Research Program (SGDX20201103095613036).

\subsubsection{\discintname}
The authors have no competing interests to declare that are relevant to the content of this article.
\end{credits}

\bibliographystyle{splncs04}
\bibliography{mybibliography}

\end{document}